\documentstyle{llncs}

\title{Efficient Algorithms for Dualizing Large-Scale Hypergraphs}
\author{Keisuke Murakami\inst{1} and Takeaki Uno\inst{1}}
\institute{National Institute of Informatics, 2-1-2 Hitotsubashi, Chiyoda-ku, Tokyo 101-8430, JAPAN, 
{\tt e-mail: \{murakami,uno\}@nii.ac.jp}}

\begin{document}

\maketitle

\begin{abstract}

A hypergraph ${\cal F}$ is a set family defined on vertex set $V$.
The dual of ${\cal F}$ is the set of minimal subsets $H$ of $V$ such that
 $F\cap H \ne \emptyset$ for any $F\in {\cal F}$.
The computation of the dual is equivalent to many problems, such as
 minimal hitting set enumeration of a subset family, minimal set cover
 enumeration, and the enumeration of hypergraph transversals.
Although many algorithms have been proposed for solving the problem,
 to the best of our knowledge, none of them can work on
 large-scale input with a large number of output minimal hitting sets.
This paper focuses on developing time- and space-efficient algorithms
 for solving the problem.
We propose two new algorithms with new search methods, 
 new pruning methods, and fast techniques for the minimality check.
The computational experiments show that our algorithms are quite fast
 even for large-scale input for which existing algorithms do not
 terminate in a practical time.
\end{abstract}

\section{Introduction}

A {\em hypergraph} ${\cal F}$ is a subset family defined on a vertex set $V$,
 that is, each element (called {\em hyperedge}) $F$ of $\cal F$ is a subset
 of $V$.
The hypergraph is a generalization of a graph so that edges can have more
 than two vertices.
A {\em hitting set} is a subset $H$ of $V$ such that $H\cap F\ne \emptyset$
 for any hyperedge $F \in {\cal F}$.
 A hitting set is called {\em minimal} if it includes no other hitting set.
The {\em dual} of a hypergraph is the set of all minimal hitting sets.
The {\em dualization} of a hypergraph is to construct the dual of a
 given hypergraph.

Dualization is a fundamental problem in computer science, especially in
 machine learning, data mining, and optimization, etc.
It is equivalent to (1) the minimal hitting set enumeration of given
 subset family,
 (2) minimal set cover enumeration of given set family,
 (3) enumeration of hypergraph transversal,
 (4) enumeration of minimal subsets that are not included in any of the
  given set family, etc.
One of the research goals is to clarify the existence of a polynomial
 time algorithm for solving the problem.
The size of dual can be exponential in the input hypergraph, thus the
 polynomial time algorithm for dualization usually means an algorithm
 running in time polynomial to the input size and the output size.
Although Kachian et al.\cite{FrKc96} developed a quasi-polynomial
 time algorithm which runs in $O(N^{\log N})$ time, where $N$ is the
 input size plus output size, the existence of a polynomial time algorithm
 is still an open question.

From the importance of dualization in its application areas, a lot of
 research has aimed at algorithms that terminate in a short time on real
 world data.
The size of the dual can be exponential, but in practice, it is huge but
 not intractable.
Thus, practically efficient algorithms aim to take a short time for
 each minimal hitting set.
Reduction of the search space was studied as a way to cope with this problem
 \cite{DgLj05,FrKc96,HcBa07,KvSt99,IBE,IBE2}.
Finding a minimal hitting set is easy; one removes vertices one by one
 unless each has an empty intersection with some hyperedges.
However, finding exactly all minimal hitting sets is not easy; we have
 to check a great many vertex subsets that can be minimal hitting sets.
The past studies have succeeded in reducing the search space,
 but the computational cost was substantial, hence the current algorithms
 may take a long time when the size of the dual is large.

In this paper, we focus on developing an efficient computation for the case
 of large-scale input data with a large number of minimal hitting sets.
We looked at the disadvantages of the existing methods and
 devised new algorithms to eliminate them.
 
\begin{itemize}
\item {\em breadth-first search:} A popular search method for dualization is
 hill climbing such that the algorithm starts from the emptyset, 
and recursively adds vertices one by one until it reaches minimal hitting sets.
 The minimal hitting sets already found are stored in memory and used to
 check the minimality.
This minimality check is popular, but its memory usage is so inefficient
 so that we cannot solve a problem with many minimal hitting sets.
We alleviate this disadvantage by using a depth-first search algorithm
 with the use of the new minimality check algorithm explained below.
The algorithm proposed in \cite{KvSt99,KvSt05} uses a depth-first search,
 but its minimality check takes a long time on large hypergraphs.

\item {\em minimality check:} The time for the minimality check in a
 breadth-first search is short when the hitting sets to be checked are
 small on average, but will be long for larger hitting sets (such as size
 20 or larger).
We alleviate this disadvantage by using a new algorithm
 that does not need the hitting sets that have already been found.
We introduce a new concept, called the {\em critical hyperedge}, 
 that characterizes the minimality of hitting sets.
Computing and updating critical hyperedges can be done in a short time,
 thus we can efficiently check the minimality in a short time.

\item {\em pruning:} Several algorithms use pruning methods to reduce the
 search space, but our experiments show that these pruning methods are not
 sufficient.
We propose a simple but efficient pruning method.
We introduce a lexicographic depth-first search, and thereby remove vertices
 that can never be used and prune branches without necessary vertices.
The pruning drastically reduces the computation time.

\item {\em sophisticated use of simple data structures:}
Not many studies have mentioned the data structures or how to use them
 efficiently, despite this being a very important consideration to reduce
 the computation time.
We use both the adjacency matrix (characteristic vectors of hyperedges)
 and doubly linked lists to speed up the operations of taking intersections
 and set differences.
This accelerates the computation time in extremely sparse, extremely
 dense (use complement as input), non-small minimal hitting sets (over 10
 vertices) cases.
\end{itemize}

The paper is organized as follows.
In the following subsections, we explain the related work and related problems.
Section 2 is for preliminaries, and Section 3 describes the existing
 algorithms.
We describe our new algorithms in Section 4 and show the results of
 computational experiments in Section 5.
We conclude the paper in Section 6.

\subsection{Related Work}

There have been several studies on the dualization problem, 
 of which we shall briefly review the DL, BMR, KS and HBC algorithms.
These algorithms are classified into two types according to their structure;
 improved versions of the Berge algorithm\cite{Ber89}, and hill-climbing
 algorithms.
The Berge algorithm updates the set of minimal hitting sets iteratively, 
 by adding hyperedges one by one to the current partial hypergraph.
DL, BMR and KS are the algorithms of this type, and HBC is the hill-climbing
 type.
The candidates for minimal hitting sets are generated by gathering
 vertices one-by-one until a minimality condition is violated.
When a candidate becomes a hitting set, it is a minimal hitting set.
The HBC algorithm does this operation in a breadth-first manner.
 
The DL algorithm, proposed by Dong and Li \cite{DgLj05},
 is a border-differential algorithm for data mining.
The main difference from the Berge algorithm is that it avoids generating
 non-minimal hitting sets by increasing the problem size incrementally.
The DL algorithm starts from an empty hypergraph and adds a hyperedge
 iteratively while updating the set of minimal hitting sets.
The sizes of the intermediate sets of minimal hitting sets are likely
 smaller than that of the original hypergraph, thus we can expect that
 there will be no combinatorial explosion.
Experiments on two small UCI datasets \cite{UCI} have shown that the
 DL algorithm is much faster than their previous algorithm and the
 level-wise hill climbing algorithm.

In general, the Berge algorithm and DL algorithm are very useful
 when the hypergraph has few hyperedges, but for large hypergraphs,
 it may take a long time because of many updates.
The BMR algorithm, proposed by Bailey et al. \cite{BMR03}, starts from
 a hypergraph with few vertices with hyperedges restricted to the vertex set
 (the vertices not in the current vertex set are removed from the hyperedges).
The hyperedges grow as the vertex set increases.
The BMR algorithm first uses the Berge algorithm to solve the problem of
 the initial hypergraph, and then it updates the minimal hitting sets.
Note that Hagen tested a version of the DL algorithm instead of the
 Berge algorithm \cite{Hag08}.

Kavvadias and Stavropoulos's algorithm (KS algorithm) \cite{KvSt99,KvSt05}
 embodies two ideas; unifying the nodes contained in the same hyperedges
 and depth-first search.
These ideas help to reduce the number of intermediate hitting sets and
 memory usage.
To perform a depth-first search, they use a minimality check algorithm
 that does not need other hitting sets;
 check whether the removal of each vertex results a hitting set or not.
The KS algorithm uses an efficient algorithm for this task.

Hebert et al. proposed a level-wise algorithm (HBC algorithm) \cite{HcBa07}.
Their algorithm is a hill climbing algorithm which starts from the empty set
 and adds vertices one by one.
It searches the vertex subsets satisfying a necessary condition to be
 a minimal hitting set, called a ``Galois connection''.
A vertex subset satisfies the Galois connection if the removal of any of
 its vertexes decreases the number of hyperedges intersecting with it.
The sets satisfying the Galois connection form a set system satisfying
 the monotone property (independent set system), thus we can perform
 a breadth-first search in the usual way.
 

\subsection{Related Problems}

Dualization has many equivalent problems.
We show some of them below.\\

{\em (1) minimal set cover enumeration}\\
For a subset family $\cal F$ defined on a set $E$, a set cover $S$ is a
 subset of $\cal F$ such that the union of the members of $S$ is equal to
 $E$, i.e., $E = \bigcup_{X\in S} X$.
A set cover is called minimal if it is included in no other set cover.
We consider $\cal F$ to be a vertex set, and ${\cal F}(v)$ to be a hyperedge
 where ${\cal F}(v)$ is the set of $F\in {\cal F}$ that include $v$.
Then, for the hyperedge set (set family) $F = \{{\cal F}(v) | v\in E\}$, 
 a hitting set of $F$ is a set cover of $\cal F$, and vice versa.
Thus, enumerating minimal set covers is equivalent to dualization.\\

{\em (2) minimal uncovered set enumeration}\\
For a subset family $\cal F$ defined on a set $E$, an uncovered set $S$
 is a subset of $E$ such that $S$ is not included in any member of $\cal F$.
Let $\bar{\cal F}$ be the complement of $\cal F$, which is the set of the
 complement of members in $\cal F$, i.e.,
 $\bar{\cal F} = \{E\setminus X | X\in {\cal F}\}$.
$S$ is not included in $X\in {\cal F}$ if and only if $S$ and $E\setminus X$
 have a non-empty intersection.
An uncovered set of $\cal F$ is a hitting set of $\bar{\cal F}$, and vice
 versa, thus the minimal uncovered set enumeration is equivalent to the
 minimal hitting set enumeration.\\

{\em (3) circuit enumeration for independent system}\\
A subset family $\cal F$ defined on $E$ is called an independent system
 if for each member $X$ of $\cal F$, any of its subsets is also a member
 of $\cal F$.
A subset of $E$ is called independent if it is a member of $\cal F$,
 and dependent otherwise.
A circuit is a minimal dependent set, i.e., a dependent set which properly
 contains no other dependent set.
When an independent system is given by the set of maximal independent sets
 of $\cal F$, then the enumeration of circuits of $\cal F$ is equivalent
 to the enumeration of uncovered sets of $\cal F$.\\

{\em (4) Computing negative border from positive border}\\
A function is called Boolean if it maps subsets in $2^V$ to $\{0, 1\}$.
A Boolean function $B$ is called monotone (resp., anti-monotone)
 if it for any set $X$ with $B(X)=0$ (resp., $B(X)=1$), any subset $X'$ of
 $X$ satisfies $B(X')=0$ (resp., $B(X')=1$).
For a monotone function $B$, a subset $X$ is called a positive border if
 $B(X)=0$ and no its proper superset $Y$ satisfies $B(Y)=0$ , and is called
 a negative border if $B(X)=1$ and no its proper subset $Y$ satisfies $B(Y)=1$.
When we are given a Boolean function by the set of positive borders,
 the problem is to enumerate all of its negative borders.
This problem is equivalent to dualization, since the problem is equivalent
 to uncovered set enumeration.\\

{\em (5) DNF to CNS transformation}\\
DNF is a formula whose clauses are composed of literals connected by ``or''
 and whose clauses are connected by ``and''.
CNF is a formula whose clauses are composed of literals connected by ``and'',
 and whose clauses are connected by ``or''.
Any formula can be represented as a DNF formula and a CNF formula.
Let $D$ be a DNF formula composed of variables $x_1,...,x_n$ and clauses
 $C_1,...,C_m$.
A DNF/CNF is called monotone if no clause contains a literal with ``not''.
Then, $S$ is a hitting set of the clauses of $D$ if and only if the
 assignment obtained by setting the literals in $S$ to true gives a true
 assignment of $D$.
Let $H$ be a minimal CNF formula equivalent to $D$.
$H$ has to include any minimal hitting set of $D$ as its clause, since
 any clause of $H$ has to contain at least one literal of any clause of $D$.
Thus, a minimal CNF equivalent to $D$ has to include all minimal hitting
 sets of $D$.
For the same reason, computing the minimal DNF from a CNF is equivalent
 to dualization.

\section{Preliminaries}

A {\em hypergraph} ${\cal F}$ is a subset family $\{F_1,\ldots,F_m\}$
 defined on a vertex set $V$, that is, each element (called {\em hyperedge})
 $F$ of $\cal F$ is a subset of $V$.
The hypergraph is a generalization of a graph so that edges can contain more
 than two vertices.
A subset $H$ of $V$ is called a {\em vertex subset}.
A {\em hitting set} is a vertex subset $H$
 such that $H\cap F\ne \emptyset$ for any hyperedge $F \in {\cal F}$.
A hitting set is called {\em minimal} if it includes no other hitting set.
The {\em dual} of a hypergraph is the hypergraph whose hyperedge set is the
 set of all minimal hitting sets, and it is denoted by $dual({\cal F})$.
For example, when $V=\{1,2,3,4\}, {\cal F}=\{\{1,2\}, \{1,3\}, \{2,3,4\} \},$
 $\{1,3,4\}$ is a hitting set but not minimal, and $\{2,3\}$ is a minimal
 hitting set.
$dual({\cal F})$ is $\{\{1,2\}, \{1,3\}, \{1,4\}, \{2,3\} \}$.
It is known that ${\cal F} = dual(dual({\cal F}))$ if no $F,F'\in {\cal F}$
 satisfy $F\subset F'$.
The {\em dualization} of a hypergraph is to construct the dual of the
 given hypergraph.

$|{\cal F}|$ denotes the number of hyperedges in $\cal F$, that is $m$,
 and $||{\cal F}||$ denotes the sum of the sizes of hyperedges in $\cal F$,
 respectively.
In particular, $||{\cal F}||$ is called the {\em size} of $\cal F$.
${\cal F}_i$ denotes the hypergraph composed of hyperedges
 $\{F_1,\ldots,F_i\}$.
For $v\in V$, let ${\cal F}(v)$ be the set of hyperedges in $\cal F$ that
 includes $v$, i.e., ${\cal F}(v) = \{F | F\in {\cal F}, v\in F\}$.
For vertex subset $S$ and vertex $v$, we respectively denote $S\cup \{v\}$
 and $S\setminus \{e\}$ by $S\cup v$ and $S\setminus e$.

We introduce the new concept {\em critical hyperedge} in the following.
For a vertex subset $S\subseteq V$, $uncov(S)$ denotes the set of
 hyperedges that do not intersect with $S$, i.e.,
 $uncov(S) = \{F | F\in {\cal F}, F\cap S = \emptyset\}$.
$S$ is a hitting set if and only if $uncov(S) = \emptyset$.
For a vertex $v\in S$, a hyperedge $F\in {\cal F}$ is said to be
 {\em critical} for $v$ if $S\cap F = \{v\}$.
We denote the set of all critical hyperedges for $v$ by $crit(v,S)$, 
 i.e., $crit(v,S) = \{F | F\in {\cal F}, S\cap F = \{v\}\}$.
Suppose that $S$ is a hitting set.
If $v$ has no critical hyperedge, every $F\in {\cal F}$ includes a vertex
 in $S$ other than $v$, thus $S\setminus v$ is also a hitting set.
Therefore, we have the following property.

\begin{property}\label{crit}
$S$ is a minimal hitting set if and only if $uncov(S) = \emptyset$,
 and $crit(v,S)\ne \emptyset$ holds for any $v\in S$.
\end{property}

If $crit(v,S) \ne \emptyset$ for any $v\in S$, we say that $S$ satisfies the {\em minimality condition}.
Our algorithm updates $crit$ to check the minimality condition quickly,
 by utilizing the following lemmas.
Let us consider an example of $crit$.
Suppose that ${\cal F}=\{\{1,2\}, \{1,3\}, \{2,3,4\} \}$,
 and the hitting set $S$ is $\{1,3,4\}$.
We can see that $crit(1,S) = \{\{1,2\} \}, crit(3,S) = \emptyset,
 crit(4,S) = \emptyset$, thus $S$ is not minimal, and we can remove either
 $3$ or $4$.
For $S'=\{1,3\}$, $crit(1,S') = \{\{1,2\}\} , crit(3,S') = \{\{2,3,4\}\}$,
 thus $S'$ is a minimal hitting set.
The following lemmas are the keys to our algorithms.

\begin{lemma}\label{up}
For any vertex subset $S$, $v\in S$ and $v'\not \in S$,
 $crit(v,S\cup v') = crit(v,S) \setminus {\cal F}(v')$.
Particularly, $crit(v,S\cup v') \subseteq crit(v,S)$ holds.
\end{lemma}

\proof
For any $F\in crit(v, S)$, $(S\cup v') \cap F = \{v\}$ holds if $F$ is
 not in ${\cal F}(v')$, and thus it is included in
 $crit(v, S\cup v') \setminus {\cal F}(v)$.
Conversely, $(S\cup v') \cap F = \{v\}$ holds for any $F\in crit(v,S\cup v')$.
This means that $S \cap F = \{v\}$, and $F\in crit(v,S)$.
\qed

\begin{lemma}\label{up2}
For any vertex subset $S$ and $v'\not \in S$,
 $crit(v',S\cup v') = uncov(S)\cap {\cal F}(v')$.
\end{lemma}

\proof
Since any hyperedge not in $uncov(S)$ has a non-empty intersection with $S$,
 $F$ can never be a critical hyperedge for $v'$.
Any critical hyperedge for $v'$ includes $v'$ thus, we can see that
 $crit(v',S)\subseteq uncov(S)\cap {\cal F}(v')$.
Conversely, for any hyperedge $F$ included in $uncov(S)\cap {\cal F}(v')$,
 $F\cap (S\cup v') = \{v'\}$, thereby
 $uncov(S)\cap {\cal F}(v')\subseteq crit(v',S)$.
Hence, the lemma holds.
\qed

The next two lemmas follow directly from the above.

\begin{lemma}\label{mono}\cite{HcBa07}
If a vertex subset $S$ satisfies the minimality condition, any of its subsets also satisfy the minimality condition,
 i.e., the minimality condition satisfies the monotone property.
\end{lemma}

\begin{lemma}\label{min-cond}\cite{HcBa07}
If a vertex subset $S$ does not satisfy the minimality condition, 
 $S$ is not included in any minimal hitting set.
In particular, any minimal hitting set $S$ is maximal in the set system
 composed of vertex subsets satisfying the minimality condition.
\end{lemma}

\begin{lemma}\label{crit-size}
For any vertex subset $S$, $\sum_{v\in S} |crit(v,S)| \le |{\cal F}|$.
\end{lemma}

\proof
From the definition of the critical hyperedge, any hyperedge $F\in {\cal F}$
 can be critical for at most one vertex.
Thus, the lemma holds.
\qed

\section{Existing Algorithms}

This section is devoted to explaining the framework of the existing algorithms
 related to our algorithms: DL algorithm, KS algorithm, and HBC algorithm.
The DL algorithm starts by computing $dual({\cal F}_1)$ and then
 iteratively computes $dual({\cal F}_i)$ from $dual({\cal F}_{i-1})$.
For any $S\in dual({\cal F}_i)$, either $S\in dual({\cal F}_{i-1})$ holds,
 or $S\setminus v \in dual({\cal F}_{i-1})$ holds for $\{v \} = S\cap F_i$.
Note that when $S\in dual({\cal F}_i)$ is not in $dual({\cal F}_{i-1})$,
 $S\cap F_i$ is composed of exactly one vertex, since $crit(v,S)$ must be
 $\{F_i\}$.
However, for any $S\in dual({\cal F}_{i-1})$, $S\in dual({\cal F}_i)$ if
 $S\cap F_i \ne \emptyset$.
When $S\cap F_i = \emptyset$, $S\cup v$ with $v\in F_i$ may be in
 $dual({\cal F}_i)$.
The algorithm is as follows.

\begin{tabbing}
ALGORITHM DL (${\cal F} = \{F_1, \ldots, F_m\}$)\\
1. ${\cal D}_0:= \{\emptyset \}$\\
2. {\bf for} $i:= 1$ {\bf to} $m$\\
3. \ \ ${\cal D}_i:= \emptyset$\\
4. \ \ {\bf for each} $S\in {\cal D}_{i-1}$ {\bf do}\\
5. \ \ \ \ {\bf if} $S\cap F_i \ne \emptyset$ {\bf then} insert $S$ to ${\cal D}_i$\\
6. \ \ \ \ {\bf else} {\bf for each} $v\in F_i$ {\bf do}\\
7. \ \ \ \ \ \ {\bf if} no $S'\in {\cal D}_{i-1}$ satisfies $S' \ne S$ and $S'\subseteq S\cup v$ {\bf then} insert $S\cup v$ to ${\cal D}_i$\\
8. \ \ \ \ {\bf end for}\\
9. \ \ {\bf end for}\\
10. {\bf end for}
\end{tabbing}

After the computation, ${\cal D}_m$ is $dual({\cal F})$.
Line 7 is for checking whether $S\cup v$ is in ${\cal D}_i$ or not by
 looking for a hitting set included in $S\cup v$.
This needs basically $O(\sum ||{\cal D}_{i-1}||)$ time and is a bottleneck
 computation of the algorithm.
This part requires all of ${\cal D}_i$ memory, thus we need to perform a
 breadth-first search.
Kavadias and Stavropoulos\cite{KvSt99,KvSt05} proposed a depth-first version
 of this algorithm.
According to the hitting sets generation rule, each hitting set in
 ${\cal F}_i$ is uniquely generated from a hitting set of ${\cal F}_{i-1}$.
Thus, starting from each hitting set in ${\cal F}_1$, we perform this
 generation rule in a depth-first manner, and visit all the minimal
 hitting sets of all ${\cal F}_i$.
The algorithm does not store each ${\cal D}_i$ in memory, and it checks
 for the minimality of $S\cup v$ by checking whether $S\cup v\setminus f$ is
 a hitting set or not for each $f\in S$.
The algorithm is as follows.

\begin{tabbing}
ALGORITHM {\bf KS} ($S$, $i$)\\
1. {\bf if} $i = m$ {\bf then} {\bf output} $S$; {\bf return}\\
2. {\bf if} $S\cap F_i \ne \emptyset$ {\bf then call} {\bf KS}($S$, $i+1$)\\
3. {\bf else} {\bf for each} $v\in F_i$ {\bf do}\\
4. \ \ {\bf for each} $u\in S$ {\bf do}\\
5. \ \ \ \ {\bf if} $S\cup v\setminus u$ is a hitting set {\bf then go to} 8.\\
6. \ \ {\bf end for}\\
7. \ \ {\bf call} {\bf KS}($S\cup v$, $i+1$)\\
8. {\bf end for}
\end{tabbing}

The bottleneck is also the minimality check on line 5 that basically needs
 to access all hyperedges in ${\cal F}_{i-1}$.

The number hitting sets that are added a vertex is $|\bigcup_{i=1}^m
 {\cal D}_i \setminus dual({\cal F}) | = O(|\bigcup_{i=1}^m {\cal D}_i |)$.
The algorithms perform the minimality check for each addition, thus roughly
 speaking, the number of minimality checks in both algorithms is
 $O(|\bigcup_{i=1}^m {\cal D}_i | \times f)$ where $f$ is the average
 size of hyperedges.
For $S\in {\cal D}_i$ and $v\in S$, $crit(v,S)$ is non-empty, 
 since $v$ always has a critical hyperedge in ${\cal F}_i$.
It implies that any subset $S$ explored by the algorithm satisfies the
 minimality condition.

The minimality check is usually one of the time-consuming parts of
 dualization algorithms.
The check whether the current vertex subset is a hitting set or not is
 also a time consuming part, but it can be done by updating $uncov$,
 thus for almost all vertex subsets to be operated on,
 its cost is much smaller than the minimality check.
Therefore, the number of minimality checks would be a good measure of
 the efficiency of the search strategy.
Here, we define the {\em search space} of an algorithm by the set of vertex
 subsets that are checked the minimality.
The size of the search space is equal to the number of executed
 minimality checks.

The cost for the minimality check increases with $|{\cal D}_i|$, 
 for the DL algorithm, and with $S$ and $||{\cal F}||$ for the KS algorithm.
Thus, the DL algorithm will be faster when the $dual({\cal F})$ is small,
 whereas the KS algorithm will be faster when $||{\cal F}||$ is small and
 $S$ is small on average.



The HBC algorithm is a kind of branch and bound algorithm.
It starts from the emptyset, and chooses elements one by one.
For each element $v$, it generates two recursive calls concerned with a choice;  add $v$ to the current vertex subset, and do not add it.
When the current vertex subset becomes a hitting set, it checks the
 minimality, and outputs it if minimal.
To speed up the computation, the algorithm prunes branches through the use
 of the so called Galois condition.
The Galois condition for $S$ and $v\not\in S$ is
 $|uncov(S)| = |uncov(S\cup v)|$, and when it holds, $S\cup v$ is
 never included in a minimal hitting set,
 thus we can terminate the recursive call with respect to $S\cup v$.
The Galois condition is equivalent to our minimality condition, since it
 is equivalent to $crit(v,S\cup v) = \emptyset$\footnote{the Galois condition
 is proposed in 2007\cite{HcBa07}, while $crit$ is proposed in
 2003\cite{IBE,IBE2}.
The term ``minimality condition'' first appears in \cite{HcBa07}.}.
The algorithm is written as follows.

\begin{tabbing}
ALGORITHM HBC (${\cal F} = \{F_1, \ldots, F_m\}$)\\
1. ${\cal D}_0:= \{\emptyset \}$ ; $i = 0$\\
2. {\bf while} ${\cal D}_i \ne \emptyset$\\
3. \ \ {\bf for each} $S\in {\cal D}_i$ {\bf do}\\
4. \ \ \ \ {\bf if} $uncov(S) = \emptyset$ {\bf then} output $S$\\
5. \ \ \ \ {\bf for each} $v$ larger than maximum vertex in $S$ {\bf do}\\
6. \ \ \ \ \ \ {\bf if} $S\cup v$ satisfies the Galois condition {\bf then} insert $S$ to ${\cal D}_i$\\
7. \ \ \ \ {\bf end for}\\
8. \ \ {\bf end for}\\
9. {\bf end while}\\
\end{tabbing}

If the pruning method is only the Galois condition, the vertex subsets
 to be explored by the algorithm is all the non-hitting sets satisfying
 the minimality condition.
Thus, the size of search space of HBC algorithm is no less than that of
 DL algorithm.
On contrary, DL and KS algorithms has to update the minimal hitting sets
 even if they do not change, thus the HBC algorithm has an advantage
 in this point.

\section{New Search Algorithms and Minimality Check}

We propose two depth-first search (branch and bound) algorithms for
 dualization problem.
The main differences from the existing algorithms are to use $crit$ for
 the minimality condition check, and pruning methods to avoid searching
 hopeless branches.
The algorithms keep lists $crit[u]$ and $uncov$ representing $crit(u,S)$
 and $uncov(S)$.
When the algorithm adds a vertex $v$ to $S$ and generates a recursive call,
 it updates $crit[]$ and $uncov$ by the following algorithm.

\begin{tabbing}
Update\_crit\_uncov ($v, crit[], uncov$)\\
1. {\bf for each} $F\in {\cal F}(v)$ {\bf do}\\
2. \ \ {\bf if} $F\in crit[u]$ for a vertex $u\in S$ {\bf then} remove $F$
 from $crit[u]$\\
3. \ \ {\bf if} $F\in uncov$ {\bf then} $uncov:= uncov\setminus F$;
 $crit[e]:= crit[e]\cup \{F\}$\\
4. {\bf end for}
\end{tabbing}

After execution, $crit[u]$ becomes $crit(u,S\cup v)$.
Since each hyperedge $F$ can be critical hyperedge for at most one vertex,
 we put $F$ on the vertex as a mark and perform step 2 in a constant time.
Thus, the time complexity of this algorithm is $O(|{\cal F}(v)|)$.
Even though this algorithm is simple, we can reduce the time complexity of
 an iteration of the KS algorithm from $O(|V|\times ||{\cal F}||)$ to
 $O(||{\cal F}||)$.

\subsection{Reverse Search Algorithm}

One of our algorithms is based on the reverse search \cite{AvFk96},
 and it can be regarded as an improved version of the KS algorithm.
Let ${\cal S} = \bigcup_{i=1}^m {\cal F}_i$,
 that is the set of vertex subsets that are operated by KS algorithm.
Let us denote the minimum $i$ such that $F_i\in crit(v,S)$ by
 $min\_crit(v,S)$, and the minimum $i$ such that $F_i\in uncov(S)$ by
 $min\_uncov(S)$.
$min\_crit(v,S)$ (resp., $min\_uncov(S)$) is defined as $m+1$ if $crit(v,S)$
 (resp., $uncov(S)$) is empty.
Using these terms, we give a characterization of $\cal S$.

\begin{lemma}\label{cals}
$S\ne \emptyset$ belongs to $\cal S$ if and only if
 $min\_crit(v,S) < min\_uncov(S)$ holds for any $v\in S$.
\end{lemma} 

\proof
Suppose that $S\in {\cal S}$, thus $S\in {\cal F}_i$ for some $i$.
We can see that $min\_uncov(S) >i$, $crit(v, S)$ includes a hyperedge
 $F_j\in {\cal F}$ with $i<j$, and thus $min\_crit(v, S) < i$ for any $v$.
Thus, $min\_crit(v, S) < min\_uncov(S)$ holds for any $v\in S$.

Conversely, suppose that $min\_crit(v, S) < min\_uncov(S)$ holds for any
 $v\in S$.
Then, we can see that $crit(v, S) \ne \emptyset$ for any $v\in S$ because
 $min\_crit(v, S) < m+1$.
Let $i = min\_uncov(S)-1$.
Note that $i\le m$.
We can then see that $S$ is a hitting set of
 ${\cal F}_i$ and $min\_crit(v, S) \le i$.
This in turn implies that $S$ is a minimal hitting set in ${\cal F}_i$,
 and thus, it belongs to $\cal S$.
\qed

For $S\in {\cal S}$, $min\_crit(S)$ is the minimum index $i$ such that $S$
 is a minimal hitting set of ${\cal F}_i$, i.e.,
 $min\_crit(S) = \max_{v\in S} \{min\_crit(v, S) \}$.
We define the {\em parent} $P(S)$ of $S$ by $S\setminus v$, where $v$ is
 the vertex such that $min\_crit(v,S) = min\_crit(S)$.
Since any $F_i$ is critical for at most one vertex,
 $min\_crit(S)$ and the parent are uniquely defined.
The parent-child relation given by this definition is acyclic, thus forms
 a tree spanning all the vertex subsets in $\cal S$ and rooted at the emptyset.
Our algorithm performs a depth-first search on this tree starting from
 the emptyset.
This kind of search strategy is called reverse search\cite{AvFk96}.

This search strategy is essentially equivalent to KS algorithm if we skip all
 redundant iteration in which we add no vertex to the current vertex subset.
In a straightforward implementation of KS algorithm, we have to iteratively
 compute the intersection of $F_i$ and the current vertex subset $S$
 until we meet the $F_i$ that does not intersect with $S$.
When $uncov(S)$ is not so large, it takes long time.
Particularly, when $uncov(S)=\emptyset$, we may spend $\Theta(||{\cal F}||)$
 time.
On contrary, in our strategy, we have only to maintain $uncov(S)$,
 that is much lighter.

The depth-first search starts from the emptyset.
When it visits a vertex subset $S$, it finds all children of $S$ iteratively
 and generates a recursive call for each child.
In this way, we can perform a depth-first search only by finding children of
 the current vertex subset.
The way to find the children is shown in the following lemma.

\begin{lemma}\label{child}
Let $S\in {\cal S}$ and $i = min\_uncov(S)$.
A vertex subset $S'$ is a child of $S$ if and only if \\
(1) $i < m+1$\\
(2) $S'=S\cup v$ for some $v\in F_i$, and \\
(3) $min\_crit(v', S') < i$ holds for any $v'\in S$.
\end{lemma}

\proof
Suppose that $S'$ is a child of $S$.
We can see that $uncov(S)$ is not empty, and thus (1) holds.
From the definition of the parent, $S$ is obtained from $S'$ by removing
 a vertex $v$ from $S'$.
From $min\_crit(S') < min\_uncov(S')$ and $uncov(S) = uncov(S')\cup
 crit(v, S')$, we obtain $min\_crit(v, S') = min\_crit(S') = min\_uncov(S)$.
This means that $F_i\in crit(v, S')$, and thus (2) holds.
This equation also implies that (3) holds.

Suppose that $S'$ is a vertex subset satisfying (1), (2) and (3).
From (2), we see that $min\_crit(v, S') = i$.
Since $uncov(S')>i$, this together with (3) implies that $S'$ satisfies
 the conditions in Lemma \ref{cals} and thereby is included in ${\cal S}$.
$min\_crit(v, S') = i$ and (3) leads to $min\_crit(S') = i$ and
 $P(S') = S'\setminus v = S$.
Note that condition (1) guarantees the existence of $F_i$ given condition (2),
 thus it is implicitly used in the proof.
\qed 

From Lemma \ref{child}, we can find all children of $S$ by adding each
 vertex $v\in F_i$ to $S$, and checking (3).
This can be done in a short time by updating $crit$.
The algorithm is as follows.

\begin{tabbing}
global variable: $crit[]$, $uncov$\\
ALGORITHM {\bf RS} ($S$)\\
1. {\bf if} $uncov = \emptyset$ {\bf then} {\bf output} $S$; {\bf return}\\
2. $i:= min \{j | F_j\in uncov\}$\\
3. {\bf for each} $v\in F_i$ {\bf do}\\
4. \ \ call Update\_crit\_uncov ($v, crit[], uncov$)\\
5. \ \ {\bf if} $\min\{t | F_t\in crit[f]\} < i$ for each $f\in S$ {\bf then}
 {\bf call} {\bf RS}($S'$)\\
6. \ \ recover the change to $crit[]$ and $uncov$ done in 4\\
7. {\bf end for}
\end{tabbing}

\begin{theorem}
Algorithm RS enumerates all minimal hitting sets in
 $O(||{\cal F}||\times |{\cal S}|)$ time and $O(||{\cal F}||)$ space.
\end{theorem}

\proof
Since the parent-child relationship induces a rooted tree spanning all
 vertex subsets in $\cal S$, the algorithm certainly enumerates all vertex
 subsets in $\cal S$.
Since any minimal hitting set is included in $\cal S$, all minimal hitting
 sets are found by the algorithm.
The update of $crit[]$ and $uncov$ is done in $O(|F(v)|)$ time,
 thus an iteration of the algorithm takes $O(||{\cal F}||)$ time.
In total, the algorithm takes $O(||{\cal F}||\times ||{\cal S}||)$ time.

The algorithm requires extra memory for storing $crit[]$ and $uncov$
 and for memorizing the hyperedges removed in step 4.
Since $crit[]$ and $uncov$ are pairwise disjoint, the total memory for
 $crit$ and $uncov$ is $O(m)$.
If a hyperedge is removed from a list, it will not be removed again in
 the deeper levels of the recursion, from the monotonicity of $crit$.
Thus, it also needs $O(m)$ memory.
The most memory is for ${\cal F}(v)$ of each $v$, and takes
 $O(||{\cal F}||)$ space.
\qed\

\noindent
{\bf pruning method}
Suppose that in an iteration we are operating on a vertex subset $S$, 
 and have confirmed that $S\cup v$ does not satisfy the minimality condition.
From Lemma \ref{mono}, we observe that $S'\cup v$ does not satisfy the
 minimality condition if $S\subseteq S'$.
This means that in the recursive call generated by the iteration with
 respect to $S$, we do not have to care about the addition of $v$,
 thus we remove $v$ from the candidate list for addition during the
 recursive call.
This condition also holds when $S\cup v$ is a minimal hitting set, 
 since no superset of a minimal hitting set satisfies the minimality condition.
We call the vertex $v$ satisfying one of these conditions {\em violating}.

We can apply this pruning method to the RS algorithm by finding all
 violating vertices before step 3 and can output all minimal hitting sets
 $S\cup v$ found in the process.
We then execute the loop from step 3 to step 7 only for non-violating
 vertices, so that we can avoid unnecessary recursive calls.

\subsection{Depth-first Search Algorithm}

This subsection described a simple hill-climbing depth-first search algorithm,
 whose search space is contained in that of the HBC algorithm.
We start from $S = \emptyset$, and add vertices to $S$ recursively unless
 the minimality condition is violated.
To avoid the duplication, we use a list of vertices $CAND$ that represents
 the vertices that can be added in the iteration.
The vertices not included in $CAND$ will not be added, even if the addition
 satisfies the minimality condition, i.e., the iteration given $S$ and
 $CAND$ enumerates all minimal hitting sets including $S$
 and included in $S\cup CAND$ by recursively generating calls.

Suppose that an iteration is given $S$ and $CAND$, and without loss of
 generality $CAND = \{v_1,\ldots,v_k\}$.
For the first vertex $v_1$, we make a recursive call with respect to
 $S\cup v_1$, with $CAND = CAND\setminus v_1$, to enumerate all minimal
 hitting sets including $S\cup v_1$.
After the termination of the recursive call, we generate a recursive call
 for $S\cup v_2$.
To avoid finding the minimal hitting sets including $v_1$, we give
 $CAND \setminus \{v_1, v_2\}$ to the recursive call.
In this way, for each vertex $v_i$, we generate a recursive call with
 $S\cup v_i$ and $CAND = \{v_{i+1},\ldots,v_k\}$.
This search strategy is common to many algorithms for enumerating members
 in a monotone set system, for example clique enumeration \cite{tomita}.
That is, its correctness has already been proved.

Next, let us describe a pruning method coming from the necessary condition
 to be a hitting set.
Suppose that an iteration is given $S$ and $CAND$, and let $F$ be a
 hyperedge in $uncov(S)$.
We can see that any minimal hitting set including $S$ has to include at
 least one vertex in $S$.
Thus, we have to generate recursive calls with respect to vertices in
 $CAND\cap F$, but do not have to do so for vertices in $CAND\setminus F$.

In the RS algorithm, we have to find all violating vertices before
 generating recursive calls.
In contrast, we can omit this step from our DFS algorithm.
Suppose that an iteration is given $S$ and $CAND$, and is going to generate
 recursive calls with respect to vertices in $v_1,\ldots,v_k \in F\cap CAND$.
Then, we first set $CAND$ to $CAND\setminus \{v_1,\ldots,v_k\}$.
If $v_k$ is not a violating vertex, we generate a recursive call for
 $S\cup v_k$, and add $v_k$ to $CAND$.
If $v_k$ is a violating vertex, we do not add $v_k$ to $CAND$.
In this way, when we generate a recursive call with respect to $S\cup v_h$,
 all violating vertices $v_j, j>h$ have already been found, thus there is
 no need to find all them at the beginning.
The algorithm is described as follows.

\begin{tabbing}
global variable: $crit[]$, $uncov$, $CAND$\\
ALGORITHM {\bf DFS} ($S$)\\
1. {\bf if} $uncov = \emptyset$ {\bf then output} $S$ ; {\bf return}\\
2. choose a hyperedge $F$ from $uncov$;\\
3. $C:= CAND\cap F$; $CAND:= CAND\setminus C$\\
4. {\bf for each} $v\in C$ {\bf do}\\
5. \ \ call Update\_crit\_uncov ($v, crit[], uncov$)\\
6. \ \ {\bf if} $crit(f, S')\ne \emptyset$ for each $f\in S$ {\bf then}
 {\bf call} {\bf DFS}($S\cup v$); $CAND:= CAND\cup v$\\
7. \ \ recover the change to $crit[]$ and $uncov$ done in 5\\
8. {\bf end for}
\end{tabbing}

Similar to the case of the RS algorithm, the computation time of an
 iteration is bounded by $O(||{\cal F}||)$.

\subsection{Implementation Issues }

This section is devoted to the computational techniques for improving
 efficiency.
Our data structure for representing hyperedges and ${\cal F}(v)$ is an
 array list in which the IDs of vertices or hyperedges are stored.
Using array list fastens the set operations with respect to ${\cal F}(v)$
 and list vertices in a hyperedge.
The data structure for $crit$ and $uncov$ is a doubly linked list.
In each iteration, we remove some hyperedge IDs from these lists and
 reinsert them after the termination of a recursive call.
A doubly linked list is a good data structure for these operations, 
 as it preserves the order of IDs in the list.

\subsection{Using the Adjacency Matrix for Set Operations}

When two subsets $S$ and $S'$ are represented by lists of their including
 elements, the set operations such as intersection and set difference need
 $O(|S| + |S'|)$ time.
However, when we have the characteristic vectors of $S$ and $S'$, we can
 do better.
The characteristic vector of $S$ is a vector whose $i$th element is one if
 and only if $i$ is included in $S$.
To take the intersection, we scan $S$ (or $S'$) with the smaller size,
 and choose the elements included in $S'$ (or $S$).
This check can be done in $O(1)$ time with using the characteristic vector
 of $S'$, thus the computation time is reduced to $O(\min\{|S|, |S'|\})$.
For computing $S\setminus S'$, we remove their intersection from $S$,
 thus the computation time is also the same.

Our algorithms take intersection of ($crit$ and $uncov$) and $F(v)$.
Updating the characteristic vectors of ($crit$ and $uncov$) uses
 $O(|{\cal F}|)$ memory and does not increase the time complexity.
The characteristic vectors of $F(v)$ for each $v$ requires a lot of memory
 to store, thus we use it only when $||{\cal F}||$ is larger than
 $n \times |{\cal F}| /64$, i.e., $\cal F$ is dense.
Note that in our experiments, all instances satisfied this condition.

\subsection{Choosing the Smallest Hyperedge}

In the DFS algorithm, we can choose arbitrary hyperedge in $uncov$ as $F$,
 for restricting the vertices to be added.
We choose a hyperedge including the smallest number of vertices which have not
 been pruned, so that the number of recursive calls generated will be small.
Counting such vertices in each hyperedge in $uncov$ may take time longer
 than the case just choosing one arbitrary, but our preliminary experiments
 showed that it reduced the computation time almost in half.




\subsection{ Pruning Only a Restricted Set of Items}

The pruning method described above can be applied to any vertex.
However, applying it to all possible vertices may take a long time compared
 with other parts of an iteration.
Sometimes it occurs that pruning takes a long time but only few branches
 are pruned.
Thus, to make the computation time stable, we prune only the vertices in
 $CAND\cap F$, which are the vertices to be added to the current solution.
This takes a time proportional to the time spent by an iteration,
 thus it never needs a long time.

\subsection{Inputting the Complement of the Hypergraph}

In some instances, ${\cal F}$ is quite dense, e.g., over 95\% of vertices
 are included in many $F\in {\cal F}$.
This occurs when the data has no clear structure and has many minimal
 hitting sets.
We can often find such instances in practice, such as in minimal infrequent
 vertex subset mining from maximal frequent vertex subsets.
In such cases, the instance itself takes up a lot of memory, and needs a
 long time to be operated on.
Here, we can reduce the computation time by using the complement.

The complement version of our algorithm inputs the complement of each
 $F\in {\cal F}$.
The operations of each iteration change so that the vertices to be added
 are vertices not in $F$, and taking difference in the $crit$ update
 changes to taking the intersection.
This substantially reduces the computation time, since we have to access
 only a small number of vertices/hyperedges.
In our experiments, we found that this idea works well for very dense datasets.

\section{Computational Experiments}

In this section, we show the results of our computational experiments
 comparing our algorithms with the existing algorithms.

\subsection{Codes and Environments}

Our algorithms are implemented in C, without any sophisticated library such
 as binary tree.
Existing algorithms are implemented in C++ by using the vector class in STL.
KS algorithm and Fredman Khachiyan algorithm
 (BEGK\cite{BeEk03,BeEk06}) are given by the authors.
All tests were performed on a 3.2 GHz Core i7-960 with a Linux operating system
 with 24GB of RAM memory.
Note that none of the implementations used multi-cores.
The codes and the instances are available at the author's Web cite
 (http://research.nii.ac.jp/~uno/dualization.html).

\subsection{Problem Instances}

We prepared several instances of problems in several categories as follows.
 The first category consists of randomly generated instances.
Each hyperedge includes a vertex $i$ with probability $p$.
The sizes and the probabilities are listed below.


The instances in the second category were generated by the dataset
 ``connect-4'' taken from the UCI Machine Learning Repository \cite{UCI}.
Connect-4 is a board game, and each row of the dataset corresponds to
 a minimal winning/losing stage of the first player, and a minimal hitting
 set of a set of winning stages is a minimal way to disturb wining/losing
 plays of the first player.
From the dataset of winning/losing stages,
 we took the first $m$ rows to make problem instances of different sizes.

The third instances are generated from the frequent itemset (pattern) mining
 problem.
An itemset is a hyperedge in our terminology.
For a set family $\cal F$ and a support threshold $\sigma$, an itemset is
 called {\em frequent} if it is included in at least $\sigma$ hyperedges, 
 and {\em infrequent} otherwise.
A frequent itemset included in no other frequent itemset is called a
 {\em maximal frequent itemset}, and an infrequent itemset including no
 other infrequent itemset is called a {\em minimal infrequent itemset}.
A minimal infrequent itemset is a minimal itemset included in no maximal
 frequent itemset, and any subset of it is included in at least one maximal
 frequent itemset.
Thus, the dual of the set of the complements of maximal frequent itemsets
 is the set of minimal infrequent itemsets.
The problem instances are generated by enumerating all maximal frequent sets
 from the datasets ``BMS-WebView-2'' and ``accidents'', taken from the
 FIMI repository \cite{FIMI}.
The profiles of the datasets are listed below.

The fourth instances are used in previous studies \cite{KvSt05,BeEk03}.\\
 $\bullet$ Matching graph (M($n$)): 
a hypergraph with n vertices (n is even) and $n/2$ hyperedges forming a
 perfect matching, that is, hyperedge $F_i$ is $\{2i-1, 2i \}$.
This instance has few hyperedges but a large number of minimal
 hitting sets $2^{n/2}$.\\
$\bullet$ Dual Matching graph (DM($n$)): it is dual(M($n$)).
It has $2^{n/2}$ hyperedges on $n$ nodes.
This instance has a large number of hyperedges but a small number of minimal
 hitting sets $n/2$.\\
$\bullet$ Threshold graph (TH($n$)): a hypergraph with n vertices
 ($n$ is even) and hyperedge set $\{\{i, j\}: 1 \le i < j \le n, j$ is even\}.
This instance has a small number of hyperedges $n^2/4$ and a small number of
 minimal hitting sets $n/2 + 1$.\\
$\bullet$ Self-Dual Threshold graph (SDTH($n$)):
The hyperedge set of SDTH($n$) is given as $\{\{ n-1, n \} \} \cup \{\{n-1\}
 \cup E\mid E\in $TH($n-2)\} \cup \{\{n \}\cup E\mid E\in dual($TH$(n-2))\}$.
SDTH($n$) has the same number of minimal hitting sets as its hyperedges,
 $(n-2)^2/4 + n/2 + 1$.\\
$\bullet$ Self-Dual Fano-Plane graph (SDFP($n$)): 
A hypergraph with n vertices and $(k-2)^2/4+k/2+1$ hyperedges,
 where $k = (n-2)/7$.
The construction starts with the set of lines
 in a Fano plane $H_0 = \{\{1, 2, 3\},$ $\{1, 5, 6\},$ $\{1, 7, 4\},$
 $\{2, 4, 5\},$ $\{2, 6, 7\},$ $\{3, 4, 6\},$ $\{3, 5, 7\}\}$.
Then we set $H = H_1 \cup H_2 \cup\cdots\cup H_k$,
 where $H_1,H_2, ..., H_k$ are $k$ disjoint copies of $H_0$.
The dual of $H$ is the hypergraph of all $7_k$ unions obtained by
 taking one hyperedge from each of $k$ copies of $H_0 (H_1, H_2, \cdots ,H_k)$.
We finally obtain SDFP($n$), which is a hypergraph of $1+7k+7k$ hyperedges.

\subsection{Differences}

Before showing the results, we discuss the difference between the algorithms
 from the viewpoint of algorithmic structures.
Basically, the search space of DL, KS, and our RS algorithms are the same.
 However, $DL$ and $KS$ check the same hitting sets many times, 
 while RS operates by one hitting set at most once.
In addition, our RS has a pruning method, thus the number of hitting sets
 generated may be decreased.
The search spaces of the HBC and DFS algorithms are basically the same,
 but DFS reduces it by using pruning methods.

For the minimality check, DL, BMR, and HBC algorithms access basically all
 members in ${\cal D}_i$.
Basically, this takes $O(\min \{2^{|S|}\log |{\cal F}|, ||{\cal D}||)$ time.
Some heuristics can reduce the time, but the reduction ratio would be limited.
In contrast, KS takes $O(|S|\times ||{\cal F}||)$ time, and $RS$ takes
 $O(|{\cal F}(v)|)$ time.
Thus, we can expect that

\begin{itemize}
\item DL, BMR, and HBC are faster when there are only a few minimal hitting
 sets ,\\
\item HBC is faster if the search space of DL is larger than the set of
 vertex subsets satisfying the minimality condition, for example, in the
 case that the sizes of minimal hitting sets are quite small\\
\item KS, RS, and DFS are faster when $|{\cal F}|$ is small,\\
\item RS is faster than KS when the sizes of minimal hitting sets are not
 small, and vertex unification (done by KS) does not work.
\end{itemize}

\subsection{Results}

Table \ref{winning} - \ref{bms2} compare the computation times.
In these tables, $|\cal F|$ represents the number of hyperedges,
 $|F|^*$ represents the average size of hyperedges, $|dual(\cal F)|$
 represents the number of minimal hitting sets and $|S|^*$ represents
 the average size of minimal hitting sets.
The computation time is in seconds.
Furthermore, ``-'' means that the computation time was more than 1000 seconds,
 and ``fail'' implies that the computation did not terminate normally 
 because of a shortage of memory or some error.

\begin{table}
\caption{Computation time on the dataset of winning stage in Connect-4}
\label{winning}
\begin{center}
\begin{tabular}[c]{| r | r | r | r | r | r | r | r | r |}
\hline
\multicolumn{1}{|c||}{$w$} & \multicolumn{1}{|c|}{100} & \multicolumn{1}{|c|}{200}& \multicolumn{1}{|c|}{400} & \multicolumn{1}{|c|}{800} & \multicolumn{1}{|c|}{1600} &
\multicolumn{1}{|c|}{3200} & \multicolumn{1}{|c|}{6400} & \multicolumn{1}{|c|}{12800} \\
\hline
\multicolumn{1}{|c||}{BEGK} &1.2&5.2&46&55&430&-&-&- \\\hline
\multicolumn{1}{|c||}{DL} &0.005&0.061&1.6&6.2&180&-&-&- \\\hline
\multicolumn{1}{|c||}{BMR} &0.006&0.044&0.52&0.67&17&710&-&- \\\hline
\multicolumn{1}{|c||}{HBC} &33&-&-&-&-&-&-&- \\\hline
\multicolumn{1}{|c||}{KS} &0.021&0.14&1.1&3.2&73&860&-&- \\\hline
\multicolumn{1}{|c||}{RS} &0.001 & 0.005 & 0.032 & 0.078 & 0.41 & 4.7 & 20 & 83\\\hline
\multicolumn{1}{|c||}{DFS} &0.001 & 0.006 & 0.021 & 0.056 & 0.27 & 2.6 & 11 & 48\\\hline
\hline
\multicolumn{1}{|c||}{$|\cal F|$}& 100 & 200 & 400 & 800 & 1600 & 3200 & 6400 &12800\\
\hline
\multicolumn{1}{|c||}{$|F|^*$}& 8 & 8 & 8 & 8 & 8 & 8 & 8 &8\\
\hline
\multicolumn{1}{|c||}{$|dual(\cal F)|$}&287&1145&6069&11675&71840& 459502 & 1277933 & 11614885\\
\hline
\multicolumn{1}{|c||}{$|S|^*$}&10.70&11.95&14.15&14.84&16.46& 17.69 & 18.67 & 20.54\\
\hline
\end{tabular}
\end{center}
\end{table}

\begin{table}
\caption{Computation time on the dataset of losing stage in Connect-4}
\label{losing}
\begin{center}
\begin{tabular}[c]{| r | r | r | r | r | r | r | r | r |}
\hline
\multicolumn{1}{|c||}{$ l$} & \multicolumn{1}{|c|}{100} & \multicolumn{1}{|c|}{200} & \multicolumn{1}{|c|}{400} & \multicolumn{1}{|c|}{800} & \multicolumn{1}{|c|}{1600} & \multicolumn{1}{|c|}{3200} & \multicolumn{1}{|c|}{6400} & \multicolumn{1}{|c|}{12800} \\\hline
\multicolumn{1}{|c||}{BEGK} &4.7&51&110&340&-&-&-&- \\\hline
\multicolumn{1}{|c||}{DL} &0.11&6.4&44&210&-&-&-&- \\\hline
\multicolumn{1}{|c||}{BMR} &0.047&2.2&5.1&16&130&-&-&- \\\hline
\multicolumn{1}{|c||}{HBC} &110&-&-&-&-&-&-&- \\\hline
\multicolumn{1}{|c||}{KS} &0.057&2.6&4.6&20&97&-&-&- \\\hline
\multicolumn{1}{|c||}{RS} &0.009 & 0.052 & 0.14 & 0.41 & 1.6 & 15 & 98 & 420\\\hline
\multicolumn{1}{|c||}{DFS} &0.006 & 0.044 & 0.09 & 0.28 & 0.94 & 12 & 40 & 180\\\hline
\hline
\multicolumn{1}{|c||}{$|\cal F|$} & 100 & 200 & 400 & 800 & 1600 & 3200 & 6400 & 12800\\\hline 
\multicolumn{1}{|c||}{$|F|^*$} & 8 & 8 & 8 & 8 & 8 & 8 & 8 & 8\\\hline 
\multicolumn{1}{|c||}{$|dual(\cal F)|$} & 2341 & 22760 & 33087 & 79632 & 212761 & 2396735 & 4707877 & 16405082\\\hline 
\multicolumn{1}{|c||}{$|S|^*$} & 11.19 & 12.43 & 13.59 & 14.62 & 15.73 & 17.06 & 17.41 & 19.09\\\hline
\end{tabular}
\end{center}
\end{table}

\begin{table}
\begin{minipage}{177pt}
\begin{center}
\caption{Computation time on matching graphs}
\label{matching}
\begin{tabular}[c]{| r | r | r | r | r | r | r |}
\hline
\multicolumn{1}{|c||}{$M$} & \multicolumn{1}{|c|}{20} & \multicolumn{1}{|c|}{24} & \multicolumn{1}{|c|}{28} & \multicolumn{1}{|c|}{32} & \multicolumn{1}{|c|}{36} & \multicolumn{1}{|c|}{40} \\\hline
\multicolumn{1}{|c||}{BEGK} &0.045&0.72&1.1&4.4&36&fail \\\hline
\multicolumn{1}{|c||}{DL} &0.003&0.012&0.04&0.21&0.89&3.9 \\\hline
\multicolumn{1}{|c||}{BMR} &0.003&0.016&0.045&0.19&1.2&5.3 \\\hline
\multicolumn{1}{|c||}{HBC} &0.17&2.4&37&520&-&- \\\hline
\multicolumn{1}{|c||}{KS} &0&0.003&0.01&0.044&0.2&0.87 \\\hline
\multicolumn{1}{|c||}{RS} &0 & 0.004 & 0.013 & 0.059 & 0.25 & 1.1\\\hline
\multicolumn{1}{|c||}{DFS} &0.002 & 0.006 & 0.023 & 0.06 & 0.26 & 1.1\\\hline
\hline
\multicolumn{1}{|c||}{$|\cal F|$} &10&12&14&16&18&20\\\hline
\multicolumn{1}{|c||}{$|F|^*$} &2&2&2&2&2&2\\\hline
\multicolumn{1}{|c||}{$|dual(\cal F)|$} &$2^{10}$&$2^{12}$&$2^{14}$&$2^{16}$&$2^{18}$&$2^{20}$\\\hline
\multicolumn{1}{|c||}{$|S|^*$} &10&12&14&16&18&20\\\hline
\end{tabular}
\end{center}
\end{minipage}
\begin{minipage}{167pt}
\begin{center}
\caption{Computation time on dual matching graphs}
\label{d_matching}
\begin{tabular}[c]{| r | r | r | r | r | r | r |}
\hline
\multicolumn{1}{|c||}{$DM$} & \multicolumn{1}{|c|}{20} & \multicolumn{1}{|c|}{24} & \multicolumn{1}{|c|}{28} & \multicolumn{1}{|c|}{32} & \multicolumn{1}{|c|}{36} & \multicolumn{1}{|c|}{40} \\\hline
\multicolumn{1}{|c||}{BEGK} & 1.4 & 3.1 & 8.9 & 67 & fail & fail\\\hline
\multicolumn{1}{|c||}{DL} & 0.01 & 0.054 & 0.25 & 1.2 & 7.1 & 70\\\hline
\multicolumn{1}{|c||}{BMR} & 0.038 & 0.4 & 4.2 & 49 & 540 & -\\\hline
\multicolumn{1}{|c||}{HBC} & 0.21 & 3.3 & 57 & 900 & - & -\\\hline
\multicolumn{1}{|c||}{KS} & 0.012 & 0.071 & 0.56 & 5.6 & 60 & 780\\\hline
\multicolumn{1}{|c||}{RS} &0.007 & 0.054 & 0.5 & 4.8 & 50 & -\\\hline
\multicolumn{1}{|c||}{DFS} &0.014 & 0.075 & 0.64 & 6.8 & 73 & -\\\hline\hline
\multicolumn{1}{|c||}{$|\cal F|$} &$2^{10}$&$2^{12}$&$2^{14}$&$2^{16}$&$2^{18}$&$2^{20}$\\\hline
\multicolumn{1}{|c||}{$|F|^*$} & 10 & 12 & 14 & 16 & 18 & 20\\\hline
\multicolumn{1}{|c||}{$|dual(\cal F)|$} & 10 & 12 & 14 & 16 & 18 & 20\\\hline
\multicolumn{1}{|c||}{$|S|^*$} & 2 & 2 & 2 & 2 & 2 & 2\\\hline
\end{tabular}
\end{center}
\end{minipage}
\end{table} 

\begin{table}
\begin{minipage}{172pt}
\begin{center}
\caption{Computation time on threshold graphs}
\label{threshold}
\begin{tabular}[c]{| r | r | r | r | r | r |}
\hline
\multicolumn{1}{|c||}{$TH$} & \multicolumn{1}{|c|}{40} & \multicolumn{1}{|c|}{80} & \multicolumn{1}{|c|}{120} & \multicolumn{1}{|c|}{160} & \multicolumn{1}{|c|}{200}\\\hline
\multicolumn{1}{|c||}{BEGK} & 0.28 & 0.84 & 2.7 & 7.5 & 19 \\\hline
\multicolumn{1}{|c||}{DL} & 0.004 & 0.027 & 0.091 & 0.24 & 0.52 \\\hline
\multicolumn{1}{|c||}{BMR} & 0.009 & 0.15 & 0.6 & 2.6 & 6.6 \\\hline
\multicolumn{1}{|c||}{HBC} & - & - & - & - & - \\\hline
\multicolumn{1}{|c||}{KS} & 0.021 & 0.34 & 2.5 & 11 & 35 \\\hline
\multicolumn{1}{|c||}{RS} & 0.001 & 0.003 & 0.016 & 0.019 & 0.048\\\hline
\multicolumn{1}{|c||}{DFS} & 0 & 0.003 & 0.01 & 0.026 & 0.037\\\hline
\hline
\multicolumn{1}{|c||}{$|\cal F|$} & 400 & 1600 & 3600 & 6400 & 10000\\\hline
\multicolumn{1}{|c||}{$|F|^*$} & 2 & 2 & 2 & 2 & 2\\\hline
\multicolumn{1}{|c||}{$|dual(\cal F)|$} & 21 & 41 & 61 & 81 & 101\\\hline
\multicolumn{1}{|c||}{$|S|^*$} & 29.05 & 59.02 & 89.02 &119.01 & 149.01 \\\hline
\end{tabular}
\end{center}
\end{minipage}
\begin{minipage}{172pt}
\begin{center}
\caption{Computation time on self-dual threshold graphs}
\label{sd_threshold}
\begin{tabular}[c]{| r| r | r | r | r | r |}
\hline
\multicolumn{1}{|c||}{$SDTH$} & \multicolumn{1}{|c|}{42} & \multicolumn{1}{|c|}{82} & \multicolumn{1}{|c|}{122} & \multicolumn{1}{|c|}{162} & \multicolumn{1}{|c|}{202}\\\hline
\multicolumn{1}{|c||}{BEGK} & 0.53 & 3.3 & 27 & 110 & 310 \\\hline
\multicolumn{1}{|c||}{DL} & 0.008 & 0.052 & 0.2 & 0.56 & 1.3 \\\hline
\multicolumn{1}{|c||}{BMR} & 0.012 & 0.19 & 0.87 & 2.7 & 7.2 \\\hline
\multicolumn{1}{|c||}{HBC} & - & - & - & - & - \\\hline
\multicolumn{1}{|c||}{KS} & 0.057 & 1 & 6.3 & 25 & 74 \\\hline
\multicolumn{1}{|c||}{RS} & 0.002 & 0.01 & 0.017 & 0.049 & 0.065\\\hline
\multicolumn{1}{|c||}{DFS} & 0.001 & 0.01 & 0.025 & 0.041 & 0.068\\\hline
\hline
\multicolumn{1}{|c||}{$|\cal F|$} & 422 & 1642 & 3662 & 6482 & 10102\\\hline
\multicolumn{1}{|c||}{$|F|^*$} & 4.34 & 4.42 & 4.45 & 4.46 & 4.47 \\\hline
\multicolumn{1}{|c||}{$|dual(\cal F)|$} & 422 & 1642 & 3662 & 6482 & 10102\\\hline
\multicolumn{1}{|c||}{$|S|^*$} & 4.34 & 4.42 & 4.45 & 4.46 & 4.47 \\\hline
\end{tabular}
\end{center}
\end{minipage}
\end{table} 

\begin{table}
\begin{minipage}[t]{165pt}
\begin{center}
\caption{Computation time on self-dual Fano-plane graphs}
\label{sd_fano}
\begin{tabular}[c]{| r | r | r | r | r | r |}
\hline
\multicolumn{1}{|c||}{$SDFP$} & \multicolumn{1}{|c|}{9} & \multicolumn{1}{|c|}{16} & \multicolumn{1}{|c|}{23} & \multicolumn{1}{|c|}{30} & \multicolumn{1}{|c|}{37} \\\hline
\multicolumn{1}{|c||}{BEGK} & 0.043 & 1.3 & 27 & 590 & - \\\hline
\multicolumn{1}{|c||}{DL} & 0 & 0.004 & 0.22 & 22 & - \\\hline
\multicolumn{1}{|c||}{BMR} & 0.001 & 0.003 & 0.11 & 3.4 & 260 \\\hline
\multicolumn{1}{|c||}{HBC} & 0 & 0.023 & 3.2 & 540 & - \\\hline
\multicolumn{1}{|c||}{KS} & 0 & 0.002 & 0.032 & 0.64 & 26 \\\hline
\multicolumn{1}{|c||}{RS} & 0 & 0.001 & 0.022 & 0.39 & 16\\\hline
\multicolumn{1}{|c||}{DFS} & 0 & 0 & 0.014 & 0.42 & 20\\\hline
\hline
\multicolumn{1}{|c||}{$|\cal F|$} & 15 & 64 & 365 & 2430 & 16843\\\hline
\multicolumn{1}{|c||}{$|F|^*$} & 3.87 & 6.27 & 9.63 & 12.89 & 15.97 \\\hline
\multicolumn{1}{|c||}{$|dual(\cal F)|$} & 15 & 64 & 365 & 2430 & 16843\\\hline
\multicolumn{1}{|c||}{$|S|^*$} & 3.87 & 6.27 & 9.63 & 12.89 & 15.97 \\\hline
\end{tabular}
\end{center}
\end{minipage}
\begin{minipage}[t]{178pt}
\begin{center}
\caption{Computation time on randomly generated instances}
\label{p}
\begin{tabular}[c]{| r | r | r | r | r |}
\hline
\multicolumn{1}{|c||}{$p $} & \multicolumn{1}{|c|}{0.9} & \multicolumn{1}{|c|}{0.8} & \multicolumn{1}{|c|}{0.7} & \multicolumn{1}{|c|}{0.6} \\\hline
\multicolumn{1}{|c||}{BEGK} & 64 & 510 & - & - \\\hline
\multicolumn{1}{|c||}{DL} & 20 & 210 & - & - \\\hline
\multicolumn{1}{|c||}{BMR} & 1.8 & 20 & 320 & - \\\hline
\multicolumn{1}{|c||}{HBC} & 0.078 & 1.9 & 33 & 680 \\\hline
\multicolumn{1}{|c||}{KS} & 3.1 & 37 & 290 & - \\\hline
\multicolumn{1}{|c||}{RS} & 0.12 & 0.87 & 6.4 & 52\\\hline
\multicolumn{1}{|c||}{DFS} & 0.093 & 0.84 & 6.1 & 52\\\hline
\multicolumn{1}{|c||}{cRS} & 0.13 & 2.6 & 29 & 300\\\hline
\multicolumn{1}{|c||}{cDFS} & 0.087 & 1.8 & 21 & 250\\\hline
\hline 
\multicolumn{1}{|c||}{$|\cal F|$} & 1000 & 1000 & 1000 & 1000\\\hline
\multicolumn{1}{|c||}{$|F|^*$} & 45.056 & 39.898 & 35.024 & 29.953\\\hline
\multicolumn{1}{|c||}{$|dual(\cal F)|$} & 30429 & 364902 & 2509943 & 16809231\\\hline
\multicolumn{1}{|c||}{$|S|^*$} & 3.75 & 4.88 & 5.94 & 7.31 \\\hline
\end{tabular}
\end{center}
\end{minipage}
\end{table} 

\begin{table}
\caption{Computation time on all maximal frequent set from ``accidents''}
\label{ac}
\begin{center}
\begin{tabular}[c]{| r | r | r | r | r | r | r | r | r |}
\hline
\multicolumn{1}{|c||}{$ ac $} & \multicolumn{1}{|c|}{200} & \multicolumn{1}{|c|}{150} & \multicolumn{1}{|c|}{130} & \multicolumn{1}{|c|}{110} & \multicolumn{1}{|c|}{90} & \multicolumn{1}{|c|}{70} & \multicolumn{1}{|c|}{50} & \multicolumn{1}{|c|}{30} \\\hline 
\multicolumn{1}{|c||}{BEGK} & 0.54 & 3.2 & 8.7 & 22 & 87 & 430 & - & - \\\hline
\multicolumn{1}{|c||}{DL} & 0.004 & 0.042 & 0.28 & 0.98 & 4.8 & 31 & 270 & - \\\hline
\multicolumn{1}{|c||}{BMR} & 0.008 & 0.041 & 0.074 & 0.17 & 2.3 & 5.7 & 21 & 140 \\\hline
\multicolumn{1}{|c||}{HBC} & 0.004 & 0.018 & 0.064 & 0.16 & 0.95 & 3.4 & 19 & 170 \\\hline
\multicolumn{1}{|c||}{KS} & fail & fail & fail & fail & fail & fail & fail & fail \\\hline
\multicolumn{1}{|c||}{RS} & 0.001 & 0.011 & 0.02 & 0.052 & 0.26 & 0.78 & 3.3 & 32\\\hline
\multicolumn{1}{|c||}{DFS} & 0.002 & 0.013 & 0.034 & 0.05 & 0.23 & 0.76 & 3.2 & 28\\\hline
\multicolumn{1}{|c||}{cRS} & 0 & 0.007 & 0.027 & 0.05 & 0.23 & 1.4 & 12 & 230\\\hline
\multicolumn{1}{|c||}{cDFS} & 0.001 & 0.005 & 0.019 & 0.051 & 0.18 & 0.95 & 8.4 & 170\\\hline
\hline
\multicolumn{1}{|c||}{$|\cal F|$} & 81 & 447 & 990 & 2000 & 4322 & 10968 & 32207 & 135439\\\hline
\multicolumn{1}{|c||}{$|F|^*$} & 57.48 & 56.34 & 72.85 & 72.23 & 326.66 & 326.08 & 325.31 & 430.39 \\\hline
\multicolumn{1}{|c||}{$|dual(\cal F)|$} & 253 & 1039 & 1916 & 3547 & 7617 & 17486 & 47137 & 185218\\\hline
\multicolumn{1}{|c||}{$|S|^*$} & 2.57 & 3.77 & 4.25 & 4.73 & 5.09 & 5.70 & 6.46 & 7.32 \\\hline
\end{tabular}
\end{center}
\end{table}

\begin{table}
\caption{Computation time on all maximal frequent set from ``BMS-WebView2''}
\label{bms2}
\begin{center}
\begin{tabular}[c]{| r | r | r | r | r | r | r | r | r | r|}
\hline
\multicolumn{1}{|c||}{$ bms2 $} & \multicolumn{1}{|c|}{800} & \multicolumn{1}{|c|}{500} & \multicolumn{1}{|c|}{400} & \multicolumn{1}{|c|}{200} & \multicolumn{1}{|c|}{100} & \multicolumn{1}{|c|}{50} & \multicolumn{1}{|c|}{30} & \multicolumn{1}{|c|}{20} & \multicolumn{1}{|c|}{10} \\\hline 
\multicolumn{1}{|c||}{BEGK} & - & - & - & - & - & - & - & - & - \\\hline
\multicolumn{1}{|c||}{DL} & 0.87 & 3.9 & 5.4 & 94 & - & - & - & - & - \\\hline
\multicolumn{1}{|c||}{BMR} & 4.7 & 18 & 20 & 110 & 380 & 1000 & - & - & - \\\hline
\multicolumn{1}{|c||}{HBC} & 0.066 & 0.2 & 0.31 & 1.1 & 3.5 & 8.8 & 23 & 37 & 87 \\\hline
\multicolumn{1}{|c||}{KS} & fail & fail & fail & fail & fail & fail & fail & fail & fail \\\hline
\multicolumn{1}{|c||}{RS} & 0.039 & 0.12 & 0.15 & 0.87 & 9.2 & 71 & 340 & 800 & -\\\hline
\multicolumn{1}{|c||}{DFS} & 0.048 & 0.089 & 0.15 & 1.1 & 13 & 92 & 400 & 950 & -\\\hline
\multicolumn{1}{|c||}{cRS} & 0.004 & 0.009 & 0.015 & 0.056 & 0.25 & 1 & 4 & 10 & 47\\\hline
\multicolumn{1}{|c||}{cDFS} & 0.003 & 0.007 & 0.012 & 0.053 & 0.25 & 1.1 & 4.4 & 12 & 62\\\hline
\hline
\multicolumn{1}{|c||}{$|\cal F|$} & 62 & 152 & 237 & 823 & 2591 & 6946 & 17315 & 30405 & 74262\\\hline
\multicolumn{1}{|c||}{$|F|^*$} & 3338.68 & 3261.89 & 3338.18 & 3337.39 & 3336.36 & 3335.91 & 3335.23 & 3334.97 & 3334.19\\\hline
\multicolumn{1}{|c||}{$|dual(\cal F)|$} & 4616 & 16991 & 15993 & 89448 & 438867 & 1289303 & 2297560 & 3064937 & 4582209\\\hline
\multicolumn{1}{|c||}{$|S|^*$} & 1.29 & 1.88 & 1.82 & 1.99 & 2.01 & 2.02 & 2.04 & 2.07 & 2.15 \\\hline
\end{tabular}
\end{center}
\end{table}

The computation time of algorithms which store minimum hitting sets, such
 as DL and BMR, depends on $|dual(\cal F)|$ and $|S|^*$.
On the other hand, the computation time of the depth-first algorithms,
 such as KS, RS and DFS, depends on $|\cal F|$ and $|F|^*$.
In particular, RS and DFS are much faster than any other algorithm in
 almost all instances, up to 10,000 times in some cases.
The exceptions are matching graphs and dual matching graphs; both are 
extreme cases of only few small minimal hitting sets that can be
 easily found, and of few small hyperedges.
Straightforward algorithms are fast for these cases.
Also, HBC, cRS and cDFS are faster when the hypergraph is dense.
BEGK is the slowest in most instances; algorithms with smaller complexity
 are not always faster.
KS algorithm embodies an idea to unify the isomorphic vertices into
 one to reduce the number of iterations, but it seems that this is 
 not so much efficient in our experiments.
In our extra experiments, such isomorphic vertices exist in only a few
 iterations, thus the improvement brought about by unifying them would be
 limited.

Note that in instance $M$, DL is not slow even though $|dual({\cal F})|$
 is very large.
This reason would be that for any $S\cap F_i = \emptyset$ for all $i$ and
 the minimality check would not be required at all.
In several instances, BMR is slower than DL, even though it is an improved
 version.
The reason would be that BMR uses up a lot of time in preprocessing.

The following Table \ref{pru} lists the average ratios of computation
 times relative to the case without pruning.
The value is the average over all instances in the categories, and smaller
 values mean more improvement.
In some cases the ratio is slightly larger than 1.0, however basically 
 the pruning works well especially for RS.
The reason that the pruning is not so efficient for DFS is that DFS already
 has a pruning method, thus the improvement is limited.

\begin{table}
\caption{Reduction ratio of computation time by pruning method}
\label{pru}
\begin{center}
\begin{tabular}[c]{| r | r | r | r | r | r | r | r | r | r | r |} 
\hline
\multicolumn{1}{|c||}{instance}  & \multicolumn{1}{|c|}{$w$} & \multicolumn{1}{|c|}{$l$} & \multicolumn{1}{|c|}{$M$} & \multicolumn{1}{|c|}{$DM$} & \multicolumn{1}{|c|}{$TH$} & \multicolumn{1}{|c|}{$SDTH$} & \multicolumn{1}{|c|}{$SDFP$} & \multicolumn{1}{|c|}{$p$} & \multicolumn{1}{|c|}{$ac$} & \multicolumn{1}{|c|}{$bms2$}\\\hline
 \multicolumn{1}{|c||}{RS (all)}  & 0.37 & 0.44 & 0.73 & 0.16 & 1.00 & 0.20 & 0.30 & 0.33 & 0.34 & 0.56\\\hline
 \multicolumn{1}{|c||}{DFS (all)}  & 0.98 & 1.09 & 1.08 & 1.03 & 0.86 & 1.03 & 0.46 & 0.94 & 0.73 & 1.01\\\hline
 \multicolumn{1}{|c||}{RS (large)}  & 0.19  & 0.19  & 0.96  & 0.11  & 0.77  & 0.12  & 0.15  & 0.29  & 0.17  & 0.33\\\hline
 \multicolumn{1}{|c||}{DFS (large)}  &0.96  & 0.95  & 1.01  & 1.00  & 0.83  & 1.19  & 0.68  & 0.94  & 0.44  & 1.00 \\\hline
\end{tabular}
\end{center}
\end{table}

We also evaluated the total memory usage of each algorithm.
The memory usage mainly depends on the number of minimal hitting sets,
 thus we display two extreme cases; dual matching graphs $DM$ and the
 randomly generated instances $p$.
In the results, all algorithms use a lot of memory when $|\cal F|$ is large.
In particular, DL, BMR and HBC use more memory, since STL library uses a
 much memory for the sake of making variable operations more efficient.
Our algorithm and KS algorithm are quite stable to increasing the number
 of minimal hitting sets, while the others are quite sensitive.
KS uses 2.3 megabytes of memory while ours use 12 megabytes.
However, 12 megabytes are used by standard library (libc),
 thus basically the difference can be ignored.

\begin{table}
\begin{minipage}[t]{165pt}
\caption{Total memory requirement for dual matching graphs (megabytes)}
\label{memdm}
\begin{center}
\begin{tabular}[c]{| r | r | r | r | r | r | r |}
\hline
\multicolumn{1}{|c||}{$DM$} & \multicolumn{1}{|c|}{20} & \multicolumn{1}{|c|}{24} & \multicolumn{1}{|c|}{28} & \multicolumn{1}{|c|}{32} & \multicolumn{1}{|c|}{36} & \multicolumn{1}{|c|}{40} \\\hline
\multicolumn{1}{|c||}{BEGK} & 51 & 51 & 58 & 65 & fail & fail\\\hline
\multicolumn{1}{|c||}{DL} & 43 & 45 & 51 & 160 & 580 & 2300\\\hline
\multicolumn{1}{|c||}{BMR} & 21 & 24 & 41 & 110 & 610 & -\\\hline
\multicolumn{1}{|c||}{HBC} & 25 & 76 & 710 & 7900 & - & -\\\hline
\multicolumn{1}{|c||}{KS} & 1.9 & 3 & 8 & 25 & 94 & 300\\\hline
\multicolumn{1}{|c||}{RS} & 13 & 13 & 15 & 24 & 66 & -\\\hline
\multicolumn{1}{|c||}{DFS} & 13 & 13 & 15 & 24 & 66 & -\\\hline
\hline
\multicolumn{1}{|c||}{$|\cal F|$} &$2^{10}$&$2^{12}$&$2^{14}$&$2^{16}$&$2^{18}$&$2^{20}$\\\hline
\multicolumn{1}{|c||}{$|F|^*$} & 10 & 12 & 14 & 16 & 18 & 20\\\hline
\multicolumn{1}{|c||}{$|dual(\cal F)|$} & 10 & 12 & 14 & 16 & 18 & 20\\\hline
\multicolumn{1}{|c||}{$|S|^*$} & 2 & 2 & 2 & 2 & 2 & 2\\\hline
\end{tabular}
\end{center}
\end{minipage}
\begin{minipage}[t]{178pt}
\caption{Total memory requirement for randomly generated instances (megabytes)}
\label{memp}
\begin{center}
\begin{tabular}[c]{| r | r | r | r | r |}
\hline
\multicolumn{1}{|c||}{$p $} & \multicolumn{1}{|c|}{0.9} & \multicolumn{1}{|c|}{0.8} & \multicolumn{1}{|c|}{0.7} & \multicolumn{1}{|c|}{0.6} \\\hline
\multicolumn{1}{|c||}{BEGK} & 51 & 130 & - & -\\\hline
\multicolumn{1}{|c||}{DL} & 49 & 120 & - & -\\\hline
\multicolumn{1}{|c||}{BMR} & 30 & 100 & 660 & -\\\hline
\multicolumn{1}{|c||}{HBC} & 23 & 63 & 850 & 13000\\\hline
\multicolumn{1}{|c||}{KS} & 2.3 & 2.3 & 2.3 & -\\\hline
\multicolumn{1}{|c||}{RS} & 12 & 12 & 12 & 12\\\hline
\multicolumn{1}{|c||}{DFS} & 12 & 12 & 12 & 12\\\hline
\multicolumn{1}{|c||}{cRS} & 12 & 12 & 12 & 12\\\hline
\multicolumn{1}{|c||}{cDFS} & 12 & 12 & 12 & 12\\\hline
\hline
\multicolumn{1}{|c||}{$|\cal F|$} & 1000 & 1000 & 1000 & 1000\\\hline
\multicolumn{1}{|c||}{$|F|^*$} & 45.056 & 39.898 & 35.024 & 29.953\\\hline
\multicolumn{1}{|c||}{$|dual(\cal F)|$} & 30429 & 364902 & 2509943 & 16809231\\\hline
\multicolumn{1}{|c||}{$|S|^*$} & 3.75 & 4.88 & 5.94 & 7.31 \\\hline
\end{tabular}
\end{center}
\end{minipage}
\end{table}

\section{Conclusion}

We proposed efficient algorithms for solving the dualization problem.
The new depth-first search type algorithms are based on reverse search
 and branch and bound with a restricted search space.
We also proposed an efficient minimality condition check method that
 exploits a new concept called ``critical hyperedges''.
Computational experiments showed that our algorithms outperform the existing
 ones in almost all cases, while using less memory even for very large-scale
 problems with up to millions of hyperedges.
In some cases, though, our algorithms take a long time for
 the minimality check.
Shortening this time will be one of the future tasks.
More efficient pruning methods are also an interesting topic of future work.

\section*{Acknowledgments}
Part of this research is supported by the Funding Program for World-Leading
 Innovative R\&D on Science and Technology, Japan.
We thank Khaled Elbassion and Elias C. Stavropoulos for providing us with
 the programs used in our experiments.

\end{document}